\documentclass[10pt, conference]{IEEEtran}
\ifCLASSINFOpdf
\else
\fi

\usepackage{multirow}
\usepackage{graphicx}

\begin{document}
%
\title{``Tell me and I forget, teach me and I may remember, involve me and I learn'': changing the approach of teaching Computer Organization}


\author{\IEEEauthorblockN{Mat\'ias Lopez-Rosenfeld}
\IEEEauthorblockA{Departamento de Computaci\'on, Facultad de Ciencias Exactas y Naturales\\
Universidad de Buenos Aires and ICC, CONICET\\
Buenos Aires, Argentina\\
mlopez@dc.uba.ar}
}


%


\maketitle

\begin{abstract}
Millennials are arriving to university sometimes uncomfortable with the methods of some courses.
Ideas that worked with previous generations of students begin to fail when digital natives receive paper and  pencil as tools.
Courses must update from old paper-based methods to hands-on and computerized versions.
The present work discusses about this update and comments on one implementation in the course Computer Organization of the Computer Science curriculum at Universidad de Buenos Aires.
It also includes some metrics that show the effectiveness of the changes in attracting and engaging the digital generation.
\end{abstract}


%
\IEEEpeerreviewmaketitle

\section*{The author}

Mat\'{i}as Lopez-Rosenfeld is a PhD student of Computer Science in Universidad de Buenos Aires. He works in an interdisciplinary field: Education, Neuroscience and Artificial Intelligence.
He got his Licenciate degree from the same University.
He was a teaching asistant (TA) from 2006 to 2013. Since 2013 he is head TA of the course Computer Organization. He also taught Computer Science in different contexts: universities of low income population and in jail to prisoners who had never had the opportunity to use a computer before.

\section{Introduction}

Computer Organization\footnote{http://www.dc.uba.ar/oc1} is one of the first courses in the Computer Science \emph{Licenciatura} curriculum\footnote{http://www.dc.uba.ar/carreras/grado/licenciatura/plan} at Universidad de Buenos Aires, Argentina.
During this course students learn how a computer works. The syllabus includes a broad array of topics including representation systems (two's complement, signed magnitude representation, ASCII chars), input/output operations, digital logic circuits, CPU architecture, memory and microarchitecture --based on Null's book and Tannenbaum's book \cite{null2014essentials,tannenbaumstructured}--. Computer Organization is the only compulsory course of the \emph{Licenciatura} in which students study hardware.

The course used to be \emph{paper-based} --all exercises and exams were completed on paper, since the syllabus was designed approximately 25 years ago, when computers were still uncommon in Argentina (1.35 personal computers per 100 people \cite{computersPerPeople}).

Nowadays almost everyone has a laptop. Especially in Argentina, where the national government has provided laptops to every high school student in the country since 2010 (over 3,000,000 laptops). When Millennials --the generation who grew up around computers and with access to Internet (since 2011 50\% of the population has access to internet and nowadays almost 70\% do\cite{internetAccess})-- arrive at university with their laptops, they prefer to keep using them instead of pen and paper, which seem outdated and tedious to them. This change in the societal context is a challenge that many academic institutions has to deal with --and it is also an opportunity to update our teaching methods--.

This article is a report about taking advantage of this opportunity. It demonstrates the result of a 4-year process of change towards a \emph{hands-on} methodology in which students are encouraged to solve problems using computers, simulators and robots.

In Argentina, high schools have different orientations: commerce, art, technical, among others. Only the students from some technical schools arrive at university with a background in digital components, electrical currents, transistors and programming languages.

When starting the Computer Organization course, the average student only knows basic concepts of logic, algebra and algorithms. Students coming from technical high schools, which account for a small proportion of the course population, have a much more solid background and find the course relatively easy.
However, in the paper-based methodology, most students did not have enough background knowledge to address the gap needed to understand how digital circuits work or design small devices.
Most of these students tried to pass the course by merely studying enough to pass the exams, without trying to think beyond that. Some of them succeeded, but most failed. The presence of students from technical schools may have helped sustain the illusion that enough students were passing fairly, and that the rest were simply not working hard enough.

Many students tend to dislike this course, perhaps because they do not see its relevance for a career in industry as a software engineer. However, some foundational knowledge of Computer Organization is important in order to become a Computer Scientist. Dr.~Adri\'an Cristal\footnote{Barcelona Supercomputing Center - CSIC (IIIA) - UPC} explained its importance to our class when invited as a guest lecturer on Computer Architecture. He said,
``You can drive a car, but you will only really understand how to get the best performance out of the car if you understand how it works. It's the same with computers --if you want to excel as a software engineer, in order to be able to get the best out of it, you need some understanding of what is going on inside.''

As a teacher, one tends to have the impression that there are different types of students: those who want to learn, those who want to pass and those who do not care about the subject at all. One major challenge of teaching is to successfully engage all types of students by providing them with the scaffolding needed to master the course content.

After a midterm exam, a student once told me: ``I didn't know anything at all about digital electronics. I just studied for the other three topics on the test. I know I'm \emph{supposed} to understand all the topics, but I also know there's no way I'll ever be able to understand digital logic. I can only afford to fail one item, so I chose to fail that one.''

Her heuristic to pass the exam disappointed and saddened me.
Obviously there were some students who did not enjoy an important part of the course.
A change was needed, in other words, the way in which we taught Computer Organization had to be updated. We had the course content and the technology to move from a paper-based  course to a hands-on approach. In this report, I discuss the changes to the course material and design which were deemed necessary  to make the course both more enjoyable and more challenging. The following sections provide details on the proposed changes, their expected outcome and some of the results that were observed after implementing them.

\section{Proposed Changes}

In the original course design, the professor gave theory-oriented lectures that presented the topics from a formal perspective, and a teaching assistant (TA) solved exercises on the blackboard during the practical-oriented part of the course. These exercises were intended to prepare students for homework exercises, which were mainly designed to be solved using paper, pencil and calculator. Solving these homework exercises was not mandatory, but served as important preparation for two midterm exams which included similar exercises. Students could check their solutions with the TA. Students were required to pass both midterms in order to be eligible to take the final exam.

The course would transition from paper-based exercises to an experimental problem-solving method that aims to engage the students in the process of building a computer in an empirical manner. By adding workshops to the course, we hoped to promote problem solving abilities as well as encourage questions about and beyond the goals of the workshop. Experimenting further becomes easier in an exploration-friendly environment.

The coursework would to be better comprehended by students because each of the workshops would be designed as a step towards the construction of a computer. This scaffolded construction enables every student to test, interact with and understand the different steps involved in building the final product.

The course assessments also needed to change. Instead of multiple paper-based exams, we decided to track students' performance in the workshops. An integrative, individual final exam would be used to determine each student's grade.

In summary, we decided to transition from paper-based methods to hands-on methods, provide a motivating experience with a tangible final result, and replace the two midterm exams with continuous workshop assessments.

\section{Experience}

The changes to the course were implemented  gradually over a period of 4 years. Initially, the course comprised 8 units, each with a list of homework exercises, and 2 midterm exams consisting of 4 exercises each, covering every unit.

The modified course reduced the protagonism of homework. The newly designed homework prepares students for the unit prior to the workshop.
It only includes introductory exercises to the topic and some of the complex exercises become part of the workshops.
We recommended that students completed the exercises to be confident when using the computers during the workshop.
Each workshop lasted 4 hours --the same amount of time previously spent in classes with blackboard exercises.

In the workshops, students worked in pairs. Working with a peer helped students solve more problems on their own and resulted in concept discovery. The workshop mimics the real world in the sense that problems must be solved with co-workers.

To complete the workshops, students filled out a worksheet with the results obtained during the activity. This worksheet had checkpoints that every student pair had to validate with a TA prior to proceeding with the workshop. The TA asked questions for student consideration and encouraged further exploration of corner cases.

Table \ref{tbl:contents} illustrates homework distribution.

\begin{table*}[!t]
\caption{Course units and workload for students in terms of quantity of exercise and amount of sheets of paper.}
\label{tbl:contents}
\centering
\begin{tabular}{l|c|c|c|c|c}
\multicolumn{1}{c|}{\multirow{2}{*}{Unit}} & \multicolumn{2}{|c|}{Previous version} & \multicolumn{3}{|c}{New version} \\
& {\footnotesize\#exercises} & {\footnotesize\#sheets} & {\footnotesize\#exercises} & {\footnotesize\#sheets} & {\footnotesize\#Workshops} \\ \hline
1. Data Representation                 & 21 & 5 & 14 & 3 & - \\
2. Boolean Algebra and Digital Logic   & 27 & 9 & 18 & 4 & 2 \\
3. CPU Architecture                    & 14 & 4 &  9 & 3 & 1\\
4. Instruction Set Architecture        & 14 & 5 &  9 & 3 & -\\
5. Input/Output                        & 13 & 7 &  8 & 4 & 1\\
6. Buses                               & 18 & 5 & 11 & 4 & 1\\
7. Microarchitecture                   & 12 & 6 &  9 & 5 & 2\\
8. Cache                               & 18 & 7 &  6 & 2 & 1\\
\end{tabular}
\end{table*}

The following list is a brief summary of the workshops and their relationship with the course units and with other workshops:

\begin{figure}[!t]
\centering
\includegraphics[width=0.35\textwidth,trim={0 0 0 6.5cm},clip]{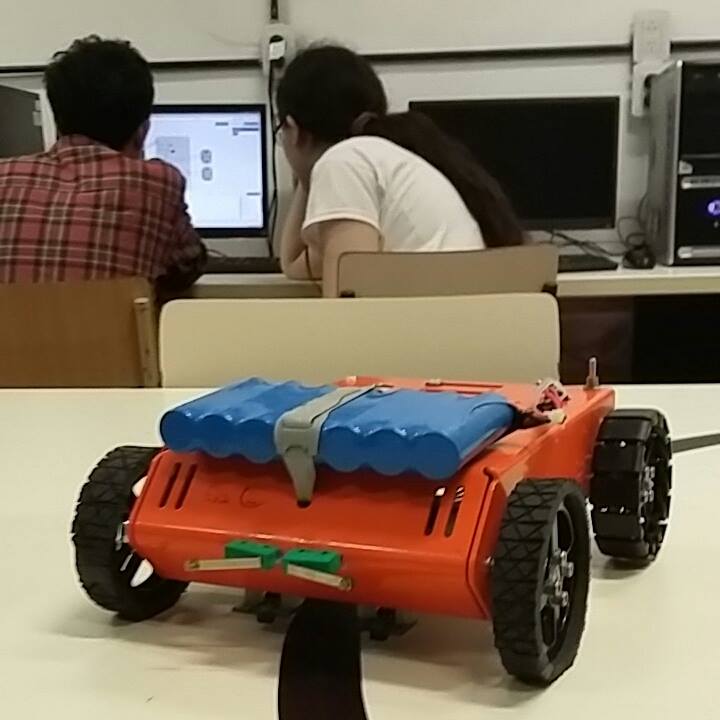}
\caption{The Input/Output workshop with the Arduino based robot-car following the line.}
\label{fig:input-output}
\end{figure}

\begin{itemize}
  \item Build an ALU using LogiSim\footnote{http://www.cburch.com/logisim/}, an educational tool for designing and simulating digital logic circuits.
  This workshop includes doing binary additions and binary subtractions on paper to validate the hardware that is built. 
  \item Build a 4-register circuit with an ALU using LogiSim.
  Move data from one register to another and make calculations using the ALU from the previous workshop.
  This workshop prepares the path to understand future units.
  \item Execute cycle simulator using a specially built simulator of an x86-type machine. Track every step of execution of a von Neumman architecture; compute offsets and labels; predict results that can be verified on the simulator. 
  \item Make a robot-car follow a line usign an Arduino-based robot. Implement routines to modify the behavior of the robot when sensing a line (see Fig.~\ref{fig:input-output}). Different protocols are presented as Finite State Machines and must be translated to code using polling or interruptions, depending on the protocol requirements. 

  \item Buses: a Python simulator of a synchronic bus provides the interface in which students write the bus controllers for master-slave protocols of various devices. An oscilloscope-like interface allows them to observe the bus and verify what their controllers are doing. 
  \item Microarchitecture 1 and 2. This longer workshop is divided in two parts. Students have a nearly complete microprogrammed computer. They must recognize control signals, the datapath, components and operation codes, and complete the machine with some logical components and microcodes. They are also provided with an assembler that translates their programs into binary code. They must code problems and run the code in this machine. They control  the entire computer. 
  \item Cache. Using another Python-based simulator, students experiment with different organizations and removal strategies and strive to achieve better performance for different memory access sequences. 
\end{itemize}

Ideally, the sequence of workshops would incrementally reach the goal of constructing every part of a computer by having each workshop reuse the tangible results that were built in the previous ones. In the current implementation, however, some of the projects actively include the concrete results of the previous workshops, while others only reuse the conceptual knowledge accrued in the previous ones. We expect this to evolve closer to our goal in the future.

Semesters at the UBA last 16 weeks. Students have a workshop approximately every 2 weeks.


\section{Reflections on the Experience}

We now present information to help determine whether the changes added value and whether students were able to better understand the course material and topics.

\begin{figure*}[!t]
\centering
\includegraphics[width=0.85\textwidth]{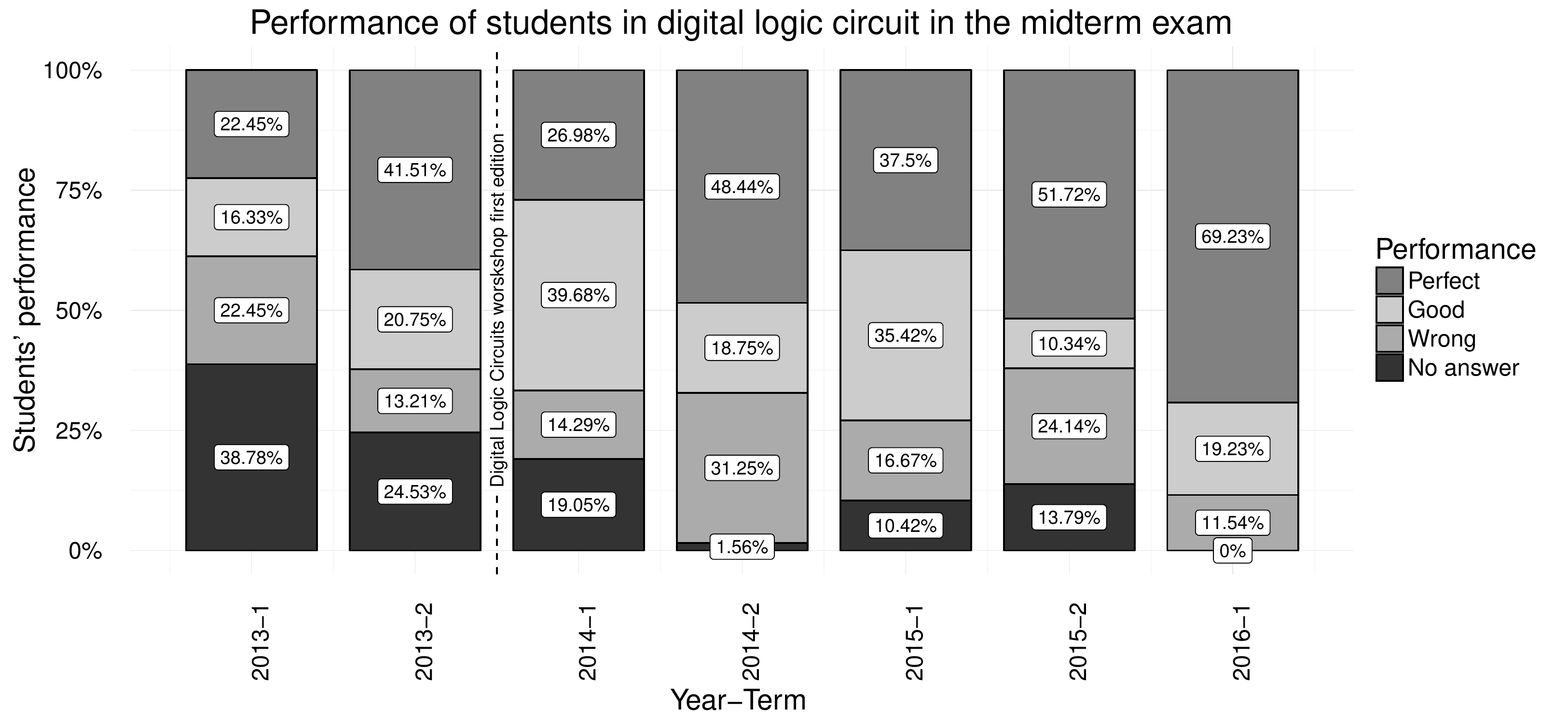}
\caption{Performance of students in digital logic circuit in the midterm exam since 2013. The grades are split in four: Perfect (when solves everything right), Good (problems in the solutions presented, such as corner cases and missing part of the assignment), Wrong (the solution presented does not solve the situation presented), and No answer (nothing at all).}
\label{fig:digital-logic}
\end{figure*}

One of the motivating concerns was that students did not seem to like digital logic, and occasionally even refused to attempt to solve exercises. This was not a problem in the new version.
Students now work extensively with digital logic components over 4 workshops, integrating this topic with the rest of the course.
Resistance to digital logic was gradually reduced over the transition. Fig. \ref{fig:digital-logic} illustrates midterm exam performance as the transition progressed. Digital logic workshops were first included in 2014. In addition to the increase in the percentage of students that solved the exercise with full marks, another interesting result is that almost all students now at least attempt to solve it. In the classroom, students appear to be more confident and comfortable with digital logic.

Additionally, students are confronted with many different assembly languages, so that they can generalize the idea and become able to use them in any dialect, ranging from x86 to MIPS and PicoBlaze. The language is no longer an obstacle.

Remarkably, in 2016, when Arduino was included as the platform for the Input/Output workshop, a group of students commented that they had bought an Arduino to \emph{be able to keep playing} at home (see Fig. \ref{fig:input-output}). This seemed like a good indicator of the success of the workshop's goal.
The first version of this workshop was based on PIC16F84A; students had to turn LEDs on and off. They did not enjoy this workshop and found it confusing. Moving a robot and programming it to follow a line, however, was very successful.

Furthermore, the hands-on course model helped students understand exactly how binary addition and subtraction operate, and in particular, what \texttt{BORROW} means. Since students had to implement binary addition and subtraction in the digital logic workshop, they all had to understand the concept. Corner cases required by the TA confirmed a genuine understanding of binary addition and subtraction. Prior the inclusion of the workshop in which students implement subtractions with a digital logic circuit, we had observed several mistakes whenever a subtraction occurred in the execution cycle workshop.

Moreover, students found the dynamic of the course enjoyable: they really like making things and understanding how they work.

Table \ref{tbl:results} shows students' grades. The correlation between the number of workshops and the percentage of students passing the course can be easily observed.

\begin{table}[!t]
\caption{Performance of the students and number of workshops per term}
\label{tbl:results}
\centering
\begin{tabular}{c|c|c|c}
Year-Term & Pass the course & \# Midterm exams & \# Workshops \\ \hline
2016 - 2 & 84.4\% & 1 & 8 \\  
2016 - 1 & 85.2\% & 1 & 7 \\  
2015 - 2 & 66.7\% & 2 & 6 \\  
2015 - 1 & 72.2\% & 2 & 5 \\  
2014 - 2 & 62.7\% & 2 & 5 \\  
2014 - 1 & 65.4\% & 2 & 4 \\  
2013 - 2 & 56\%   & 2 & 3 \\  
2013 - 1 & 50.7\% & 2 & 3 \\  
\end{tabular}
\end{table}

To understand where additional changes were needed, including changes in the direction of the course, asking many students and listening to their answers over several course iterations, as well as the different TAs who participated at different points, was of great importance.

From a teaching perspective, I think the course changes were extremely positive. Feedback from students and TAs indicates that the course is now considered both useful and interesting. While still considered a challenging course, it is now much more engaging. The curricular content has been kept on par with the previous implementation --there has been no reduction in the breadth or depth of the syllabus.

\section{Final Comments}

The change of the teaching methodology made to the Computer Organization course of the Computer Science department of the Universidad de Buenos Aires, showed a possible path to follow when updating any course of Computer Science.
Students were faced with problem-solving situations, which are similar to the ones found in software courses.
Actually, hardware is just an implementation of an algorithm, just like software, the difference is the medium.

The digital world has been the greatest social transformer in recent history. We are still getting used to it. Teaching in university is no exception. Courses --and instructors-- must find a way to involve, engage and generate enthusiasm among the new generations of digitalized students, the Millennials.

The modifications presented herein demonstrate a change in the learning process as one way of dealing with the didactic challenges in higher education and the capabilities that Millennials bring to the classroom.

We have evaluated the transition from paper-based to hands-on aproach, a success.
The students' perception and relationship with the course changed noticeably and demonstrably.
This shows evidence consistent with Benjamin Franklin's quote, ``Tell me and I forget, teach me and I may remember, involve me and I learn''.

\bibliographystyle{IEEEtran}

\end{document}